\documentclass{amsart}
\usepackage{amscd}

\newcommand{\bbC}{{\mathbb C}}
\newcommand{\bbP}{{\mathbb P}}
\newcommand{\bbR}{{\mathbb R}}
\newcommand{\bbZ}{{\mathbb Z}}

\newcommand{\bbH}{{\mathbb H}}

\newcommand{\A}{{\mathcal A}}
\newcommand{\cS}{{\mathcal S}}
\newcommand{\cO}{{\mathcal O}}

\newtheorem{lemma}{Lemma}[section]
\newtheorem{prop}[lemma]{Proposition}
\newtheorem{thm}{Theorem}

\newtheorem{cor}[lemma]{Corollary}
\newtheorem{cori}[thm]{Corollary}
\newtheorem{defn}{Definition}

\begin{document}
\title{Boundary algebras of hyperbolic monopoles}
\author{Paul Norbury}
\address{Department of Pure Mathematics\\University of
Adelaide\\Australia 5005}
\email{pnorbury@maths.adelaide.edu.au}

\begin{abstract}
    We prove the conjecture that a monopole in three-dimensional 
    anti-de Sitter space can be completely determined by its 
    ``holographic'' image on the conformal boundary two-sphere.
\end{abstract}

\maketitle

\section{Introduction}
The concept of field theories being represented by observations at the
boundary of space-time has gained much recent interest.  In
particular, the AdS-CFT correspondence proposes a relationship between
string theory on anti-de Sitter space-time and conformal field theory
on the boundary, \cite{MalLar,WitAnt}, and this generalises to produce
invariants of conformally compact Einstein manifolds with conformal
boundary \cite{GraVol}.  This paper will be concerned with a similar
issue that arises from gauge theory on three-dimensional anti-de
Sitter space.

Atiyah \cite{AtiIns,AtiMag}, first studied monopoles over hyperbolic
space $\bbH^3={\rm AdS}\sb 3$, which is three-dimensional anti-de
Sitter space with positive definite metric.  Associated to an $SU(2)$
monopole is its mass given by the norm of the Higgs field on the
conformal boundary two-sphere.  Austin and Braam \cite{ABrBou} proved
that for a half-integer mass monopole in $\bbH^3$ the holographic
image of the monopole on the conformal boundary two-sphere completely
determines the monopole, and conjectured it to be true more generally.
In this paper we will prove the conjecture for any positive real mass
monopole in $\bbH^3$.  The holography principle does not apply to
monopoles in flat space, which can be regarded as the infinite mass
limit of monopoles in anti-de Sitter space \cite{JNoZer}.  In flat
space, the holographic image on the conformal boundary two-sphere of
any two monopoles of the same charge (a topological quantity defined
below) is the same.

The main tool in this paper is an $n$-point function $\langle
P_{z_1}\dots P_{z_n}\rangle$ defined for a given monopole and any
ordered collection of points on the conformal boundary two-sphere
$\{z_1,..,z_n\}\subset S^2_{\infty}$.  Associated to the ordered
collection of points is the set of geodesics in $\bbH^3$ running from
$z_1$ to $z_2$ and from $z_2$ to $z_3$ and so on until $z_n$ to $z_1$.
We can measure an interaction between the $n$ points on the conformal
boundary by solving a scattering equation involving the monopole along
the geodesics.  The $n$-point function is a complex number assigned to
the sequence of geodesics continuously differentiable in its variables
$(z_1,..,z_n)$.  The notation $\langle P_{z_1}\dots P_{z_n}\rangle$ is
anticipating the construction of an algebra with expectation values
given by the $n$-point function.

The calculation of an $n$-point function using solutions of the
scattering equation along geodesics in $\bbH^3$ is analogous to
an approximation to the calculation of correlation functions using
path integrals appearing in the AdS/CFT correspondence, since the
stationary phase approximation reduces the computation of the
propagator to the study of the wave equation along geodesics in
$\bbH^3$.  (For a closer analogy, perhaps it would be necessary to
integrate the $n$-point functions defined in this paper over the 
moduli space of monopoles.  We will not do this here.)

The 2-point function is used to settle the open conjecture that the
holographic image of the monopole on the conformal boundary two-sphere
determines the monopole on hyperbolic space.  The $n$-point function
enables one to construct an abstract algebra freely generated by the
points of $S^2_{\infty}$, satisfying relations given in terms of the
$n$-point function.  The 3-point function is used to prove that the
generators of the algebra behave like projections, and the 4-point
function encodes the fact that the algebra is finite-dimensional.

Before describing the main results, we will define the objects of the
paper.  A hyperbolic monopole $(A,\Phi)$ is a solution of the
non-linear Bogomolny equation $d_A\Phi=*F_A$ where $A$ is a connection
defined on a trivial rank two $SU(2)$ bundle $E$ over $\bbH^3$ with
$L^2$ curvature $F_A$ and the Higgs field $\Phi:\bbH^3\rightarrow{\bf
su}(2)$ satisfies $\lim_{r\rightarrow\infty}||\Phi||=m$, the mass of
the monopole.  The charge of the monopole is defined to be the
topological degree of the map $\Phi_{\infty}:S^2_{\infty}\rightarrow
S^2_{\infty}$.  The hyperbolic metric, featured in the Hodge star $*$,
may be replaced by the Euclidean metric, giving rise to Euclidean
monopoles.  The gauge group of maps $g:\bbH^3\rightarrow SU(2)$ acts
on the equations and we identify gauge equivalent monopoles.  The
construction of an $n$-point function from a monopole is a gauge
invariant procedure.  On the conformal boundary two-sphere, a monopole
has a well-defined limit, given by a $U(1)$ connection,
\cite{JTaVor,RadSin}, which we call the holographic image of the
monopole.

There is an integrable structure underlying hyperbolic monopoles, best
seen on the complex surface of geodesics,
$\bbC\bbP^1\times\bbC\bbP^1-\bar{\Delta}$ (where
$\bar{\Delta}\subset\bbC\bbP^1\times\bbC\bbP^1$ is the anti-diagonal.)  In the
Euclidean case, over its surface of geodesics $T\bbC\bbP^1$, twistor space
techniques are used in \cite{HitCon,HitMon} to understand the
construction of monopoles, and the conserved quantities of monopoles. 
The main tool is the scattering equation
\begin{equation}  \label{eq:scatint}
(\partial_t^A-i\Phi)s=0
\end{equation}
defined for local sections $s$ of $E$ along a geodesic in $\bbR^3$
parametrised by $t$.  In particular, those geodesics along which an
$L^2$ solution of (\ref{eq:scatint}) exists, form a compact algebraic
curve inside $T\bbC\bbP^1$, called the {\em spectral curve}.  Analogously,
solutions of (\ref{eq:scatint}) along geodesics in $\bbH^3$ are used
to study hyperbolic monopoles \cite{AtiIns,AtiMag,MSiSpe} and to
define the spectral curve of the monopole
$\Sigma\subset\bbC\bbP^1\times\bbC\bbP^1-\bar{\Delta}$.

For $z_1\neq z_2$, define the 2-point function $\langle
P_{z_1}P_{z_2}\rangle$ to be a positive real number associated to
$(A,\Phi)$ and the geodesic in $\bbH^3$ joining $z_1$ and $z_2$ on the
conformal boundary two-sphere as follows.  Along this geodesic, choose
a solution $s_+(t)$ of (\ref{eq:scatint}) that decays as
$t\rightarrow\infty$.  Notice that the parameter $t$ involves a choice
of orientation of the geodesic.  Choose a decaying solution $r_+$ of
(\ref{eq:scatint}) along the same geodesic oriented in the opposite
direction.  In terms of the parameter $t$ used for $s_+(t)$, $r_+(t)$
is a solution of the equation
\begin{equation}  \label{eq:scatint1}
(\partial_t^A+i\Phi)r=0
\end{equation}
and $r_+(t)$ decays as $t\rightarrow-\infty$.  The inner product
$(r(t),s(t))$ of any two solutions of (\ref{eq:scatint}) and
(\ref{eq:scatint1}) is independent of $t$.  If we normalise $r_+(t)$ and
$s_+(t)$ by
\begin{equation} 
\lim_{t\rightarrow\infty}\exp(mt)\| s_+\|=1,\ 
\lim_{t\rightarrow-\infty}\exp(-mt)\| r_+\|=1.
\end{equation}
then the decaying solutions are well-defined up to phase and the
number $|( r_+,s_+)|^2$ depends only on the geodesic and $(A,\Phi)$. 
Define \[\langle P_{z_1}P_{z_2}\rangle=|( r_+,s_+)|^2\] for $r_+,s_+$
defined along the geodesic joining $z_1$ and $z_2$.  The $n$-point
function is a complex number defined similarly using decaying
solutions of (\ref{eq:scatint}) along the set of geodesics running
between consecutive points of an ordered $n$-tuple of points in
$S^2_{\infty}$.  For the definition of the $n$-point function and
justification of parts of the definition of the $2$-point function
given here see Section~\ref{sec:npoint}.

\begin{thm}  \label{th:2ptspec}
    The 2-point function uniquely determines the spectral curve of
    $(A,\Phi)$.
\end{thm}

The 2-point function also encodes the holographic image of the
monopole on the conformal boundary two-sphere.  The $U(1)$ connection
on the conformal boundary two-sphere is expressed with respect to a
family of gauges related to the spectral curve of the monopole.  More
explicitly, for each point $w\in S^2_{\infty}$, the 2-point function
enables one to express the $U(1)$ connection with respect to a gauge
defined over the complement of the points $\{ z_1,..,z_k\}$ that
satisfy $(w,z_i)\in\Sigma$, the spectral curve of the monopole.  Each
such gauge is determined uniquely by properties described in
Proposition~\ref{th:gauge}.  The situation is rigid enough that the
$U(1)$ connection uniquely determines the 2-point function.
\begin{thm}  \label{th:2ptinf}
    The 2-point function determines and is determined by the
    holographic image of the monopole on the conformal boundary
    two-sphere.
\end{thm}
The spectral curve determines the monopole over hyperbolic space up to
gauge equivalence.  This is a rather deep non-constructive property of
monopoles.  It uses the (non-constructive) existence of a
trivialisation of a holomorphic line bundle over the spectral curve
and sheaf cohomological constructions to retrieve the monopole.  Using 
this we are able to conclude: 
\begin{cori}   \label{th:conj}
    The holographic image of the monopole on the conformal boundary
    two-sphere determines the monopole up to gauge equivalence.
\end{cori}

The properties of the 2-point function given in
Theorem~\ref{th:2ptinf} are proven using an algebra defined abstractly
via the $n$-point functions.

An associative algebra can be studied via the values of a linear
function, which we call expectation values, defined over the algebra. 
In some cases, the structure coefficients of the algebra, with respect
to a generating set, can be retrieved from the expectation values,
thus uniquely determining the algebra.  Conversely, one may begin with
an abstract set of generators with no {\em a priori} algebra structure
and use expectation values to define the structure coefficients of the
algebra.

Consider the algebra freely generated by the points of the conformal
boundary two-sphere, where we notate the generators by $P_z$, $z\in
S^2_{\infty}$, and add the relations 
\begin{equation}  \label{eq:rel}
    \exists c=c(z_1,z_2,..,z_n)\in\bbC,\ P_{z_1}P_{z_2}\ldots P_{z_n}
     =cP_{z_1}P_{z_n},\ {\rm when}\ \langle P_{z_1}P_{z_n}\rangle\neq 0.
\end{equation}
We suppose that the $n$-point function defined by a monopole gives 
the expectation value of the product $P_{z_1}P_{z_2}\ldots P_{z_n}$ 
and we extend this function linearly to the algebra.  Then by taking 
the expectation values of each side of (\ref{eq:rel}) we can 
calculate the scalar $c$.  This essentially defines the algebra 
structure.  

The boundary algebra of a monopole is a slight modification of the
construction of the previous paragraph.  We will add further relations
to the algebra in the form of ``non-degeneracy'' conditions, and
enlarge the algebra using derivations.

\begin{defn}   \label{def:algebra}
    Define the {\bf boundary algebra}
    \[\cS(A,\Phi)=\{\A,*,P_z\in\A,z\in S^2_{\infty},\langle\dots\rangle\}\] 
    for any hyperbolic monopole $(A,\Phi)$, to consist of:
    \begin{enumerate}
	\item an involutive algebra $(\A,*)$ defined over $\bbC$,
	\item generators $P_z=P_z^*$, for all $z\in S^2_{\infty}$,
	\item derivations $[\partial_z,\cdot]:\A\rightarrow\A$ and
	$[\partial_{\bar{z}},\cdot]:\A\rightarrow\A$, 
	\item further generators $[\partial_z,P_z]$,
	$[\partial_{\bar{z}},P_z]$,
	$[\partial_z,[\partial_z,P_z]],\ldots$ 
	\item a linear function
	$\langle\dots\rangle:\A\rightarrow\bbC$ that restricts to the
	$n$-point function of $(A,\Phi)$ on products
	$P_{z_1}P_{z_2}..P_{z_n}$, and satisfies $\langle
	a^*\rangle=\overline{\langle a\rangle}$, $\partial_z\langle
	a\rangle=\langle[\partial_z,a]\rangle$,
    \end{enumerate}
    with the relations:
    \begin{enumerate}
	\item[6.] $\langle P_{z_1}P_{z_2}\rangle=0$ $\Rightarrow$
	$P_{z_1}P_{z_2}=0$, 
	\item[7.] $\langle aP_z\rangle=0$ for almost all
	$z\in S^2_{\infty}$ $\Rightarrow$ $a=0$,
	\item[8.] $\exists c=c(z_1,z_2,a,b)\in\bbC,\
        P_{z_1}aP_{z_2}=cP_{z_1}bP_{z_2}$ when $P_{z_1}bP_{z_2}\neq 0$.
    \end{enumerate}
    where $a,b\in\A$.
\end{defn}
Crucial properties of the algebra rely on limits of the $n$-point
function.  The geodesics in $\bbH^3$ used to define the $n$-point
function pass near to approximate locations of the monopole.  As a
geodesic moves out to infinity and away from the monopole, it feels
little effect, and thus the limit of the $n$-point function as two
consecutive points come together is the $(n-1)$-point function.  This
is used together with other properties of monopoles to prove various
features of the algebra:
\begin{itemize}
    \item one can make sense of the 1-point function as
    the constant function $\langle P_z\rangle\equiv 1$, 
    \item the 2-point function takes its values on the unit interval,
    \item $P_z^2=P_z$,
    \item $P_{z_1}\neq P_{z_2}$ for $z_1\neq z_2$,
    \item $P_z[\partial_z,P_z]=0$. 
\end{itemize}
Identities involving the 4-point function arise when trying to find a
representation of the algebra in which the expectation values of
observables are given by traces.  We have been unable to directly
prove these identities, described in the conclusion.  Instead we use
the fact that such a representation produces a holomorphic map
$S^2_{\infty}\rightarrow\bbC\bbP^k$, where $k$ is the charge of
$(A,\Phi)$.  This enables us to compare $\cS(A,\Phi)$ to a similar
algebra with a known representation.
\begin{thm}
    There exists a finite-dimensional representation of $\cS(A,\Phi)$
    in which the expectation values are given by traces.
\end{thm}
The holomorphic sphere $S^2_{\infty}\rightarrow\bbC\bbP^k$, which is
reminiscent of that arising in the work of Austin and Braam
\cite{ABrBou}, proves to be the source of many further interesting
properties.  It can be obtained without the algebra and gives an
alternative proof that the connection on the conformal boundary
two-sphere determines the monopole up to gauge equivalence. It also
uncovers further features.  Amongst these is an application of
geometric invariant theory to define the centre of a hyperbolic
monopole.  One also gets new information regarding rational maps
associated to monopoles.  Specifically, given a point at infinity,
there is a one-to-one mapping between gauge equivalence classes of
monopoles and degree $k$ based rational maps $S^2_{\infty}\rightarrow
S^2$ well-defined up to a $U(1)$ action.  It has never been understood
how the rational maps for different points at infinity are related. 
The holomorphic sphere gives such a relation.  These results 
will appear elsewhere \cite{MNoHyp}.

One can take finite-dimensional sub-algebras of $\cS(A,\Phi)$ and find
further structure.  In the conclusion we describe families of
subalgebras parametrised by the spectral curve of the monopole.  This
is particularly interesting due to the conjecture of Atiyah and Murray
\cite{AtiYan,AMuMon} that spectral curves of hyperbolic monopoles may
parametrise solutions of the Yang-Baxter equation.

{\em Acknowledgements.} The author would like to thank Peter
Bouwknegt, Michael Eastwood and Michael Murray for many useful
conversations.

\section{The $n$-point function}   \label{sec:npoint}
The function $\langle P_{z_1}..P_{z_n}\rangle$ defined on $n$-tuples
of points in $S^2_{\infty}$ is invariant under cyclic permutations of
the points (and hence behaves like a trace on the boundary algebra.)
In what follows, we first define the $n$-point function $\langle
P_{z_1}..P_{z_n}\rangle$ for $z_i\neq z_{i+1}$, $z_n\neq z_1$.  This
is a fundamental quantity in that all other values of
$\langle\dots\rangle$ are derived from it.  We use limits to remove
the restriction on the $n$-tuples $\{ z_1,...,z_n\}$.

Along any geodesic of $\bbH^3$ parametrised by $t$, the scattering
equations
\begin{equation}  \label{eq:scat}
(\partial_t^A-i\Phi)s=0,\ (\partial_t^A+i\Phi)r=0
\end{equation}
are defined for local sections $s,r$ of $E$.  Any pair of solutions has
the property that the inner product $( r(t),s(t))$ is
independent of $t$, since
\[\partial_t( r(t),s(t))=((\partial_t^A+i\Phi)r(t),
s(t))+( r(t),(\partial_t^A-i\Phi)s(t))=0.\] 
It can be shown \cite{HitMon,JNoCom} that that there are solutions $s$
and $r$ unique up to respective constants that decay like
$O(\exp(-mt))$ as $t\rightarrow\infty$, respectively like $O(\exp(mt)$
as $t\rightarrow-\infty$.  Thus two non-trivial solutions $s_+,r_+$
are uniquely determined up to phase by the conditions that
\begin{equation}  \label{eq:normal}
\lim_{t\rightarrow\infty}\exp(2mt)\| s_+\|^2=1,\ 
\lim_{t\rightarrow-\infty}\exp(-2mt)\| r_+\|^2=1.
\end{equation}

$\langle P_{z_1}\dots P_{z_n}\rangle$, $z_i\neq z_{i+1}$, 
$z_n\neq z_1$

For distinct $\{ z_1,\dots,z_n\}$, $\langle P_{z_1}\dots
P_{z_n}\rangle$ is a complex number associated to $(A,\Phi)$ and the
$n$ oriented geodesics in $\bbH^3$ traveling from $z_1$ to $z_2$, then
$z_2$ to $z_3$ and so on, until $z_n$ to $z_1$.  Notate by
$r_{12},s_{12}$ the solutions $r_+,s_+$ of (\ref{eq:scat}) along the
geodesic running from $z_1$ to $z_2$ and $r_{23},s_{23}$ the solutions
$r_+,s_+$ along the geodesic running from $z_2$ to $z_3$ and so on up
to $r_{n1},s_{n1}$.  Further, align the phases of each
$r_{i,i+1},s_{i-1,i}$ as follows.  The consecutive solutions
$s_{12}$ and $r_{23}$ have the property that they define a common
subspace of the fibre of $E$ at $z_2$ at infinity, or in other words
that \[\lim_{t\rightarrow\infty}\exp(mt)s_{12}=c\lim_{t'\rightarrow-
\infty}\exp(-mt')r_{23}\] for $c\in\bbC^*$.  Choose $r_{23}$ so that
$c=1$.  Similarly, choose a phase for $r_{i,i+1}$ using $s_{i-1,i}$
and for $r_{12}$ using $s_{n1}$.  Define
\begin{equation}   \label{eq:trprod} 
\langle P_{z_1}\dots P_{z_n}\rangle=( r_{12},s_{12})( r_{23},
s_{23})\dots( r_{n1},s_{n1})
\end{equation}
which depends only on $(A,\Phi)$ and the oriented geodesics running in
order through $z_1,z_2,\dots,z_n,z_1$.  The 2-point function defined
in the introduction can be obtained by setting $n=2$ in this
construction.  In this case the function is real valued and
independent of the orientation of the geodesic and the choice of
phases.
\begin{lemma}  \label{th:cont}
When $z_i\neq z_{i+1}$, $z_n\neq z_1$, the $n$-point function $\langle
P_{z_1}\dots P_{z_n}\rangle$ is continuously differentiable in
$z_1,..,z_n$.
\end{lemma}
\begin{proof}
Fix $z_2,z_3,\dots,z_n$ and vary $z_1=z$.  The product on the right
hand side of (\ref{eq:trprod}) defining $\langle P_zP_{z_2}\dots
P_{z_n}\rangle$ contains the $z$ dependent sections $r_{12}(z)$,
$s_{12}(z)$, $r_{n1}(z)$ and $s_{n1}(z)$ with the others constant as
$z$ varies.  In \cite{HitMon} (and \cite{JNoCom} for hyperbolic
monopoles) it was shown using a bijection between nearby solutions
that the assignment of $r_{12}(z)$, etc, is continuously
differentiable in $z$.  Thus, the same is true of inner products
involving the $z$ dependent sections, such as $\langle P_zP_{z_2}\dots
P_{z_n}\rangle$.
\end{proof}

For a general $n$-tuple of points $\{z_1,...z_n\}$, we define $\langle
P_{z_1}\dots P_{z_n}\rangle$ by continuity.  Lemma~\ref{th:cont} shows
that such a definition is consistent.  The following lemma explicitly
calculates the limits that arise when two points $z_i$ and $z_{i+1}$
come together.

\begin{lemma}  \label{th:square}
    The 2-point function satisfies $\lim_{z_1\rightarrow z_2}\langle
    P_{z_1}P_{z_2}\rangle=1$ and the $n$-point function satisfies
    $\lim_{z_1\rightarrow z_2}\langle P_{z_1}P_{z_2}P_{z_3}\dots
    P_{z_n} \rangle=\langle P_{z_2}P_{z_3}\dots P_{z_n}\rangle.$
\end{lemma}
\begin{proof}
    We will prove only $\lim_{z_1\rightarrow z_2}\langle
    P_{z_1}P_{z_2}\rangle=1$ since the proof of the limit of the
    $n$-point function is essentially the same.  We define $\langle
    P_{z_1}P_{z_2}\rangle=|( r_+,s_+)|^2$ for solutions of
    (\ref{eq:scat}) satisfying (\ref{eq:normal}).  If the connection
    is trivial and the Higgs field is constant,
    \begin{equation}  \label{eq:unpert}
	\partial_t^A\pm i\Phi=\partial_t\pm
	i\left(\begin{array}{cc}im&0\\0&-im \end{array}\right)
    \end{equation}
    then $r_+=\exp(mt)(1\ 0)$ and $s_+=\exp(-mt)(1\ 0)$ so $(
    r_+,s_+)=1$ as required.

    As $z_1\rightarrow z_2$, the connection and Higgs field become
    more trivial and constant, respectively.  More precisely, there
    exists a gauge in which
    \begin{equation}  \label{eq:pert}
	\partial_t^A\pm i\Phi=\partial_t\pm
	i\left(\begin{array}{cc}im&0\\0&-im
	\end{array}\right)+\epsilon\cdot C\exp(-m|t|)
    \end{equation}
    where $C$ is constant and $\epsilon\rightarrow 0$ as
    $z_1\rightarrow z_2$.  This follows from Rade \cite{RadSin}.
    
    Levinson's theorem \cite{CLeThe} uses a contraction mapping
    argument to show that solutions $r_+$ on $(-\infty,0]$ and $s_+$
    on $[0,\infty)$ of (\ref{eq:pert}) (using $i\Phi$ and $-i\Phi$
    respectively) are in one-to-one correspondence with solutions of
    (\ref{eq:unpert}).  Moreover, the norm of the difference between
    corresponding solutions is controlled by the $L^1$ norm of the
    perturbation term $\epsilon\cdot C\exp(-m|t|)$.

    In other words, as $z_1\rightarrow z_2$, the solutions $r_+$ and
    $s_+$ tend uniformly to the solutions of (\ref{eq:unpert}) on
    $(-\infty,0]$ and $[0,\infty)$ respectively, and in fact on any
    $(-\infty,R]$ and $[-R,\infty)$.  The inner product $( r_+,s_+)$
    can be calculated at any point $t\in\bbR$, in particular
    $t\in[-R,R]$ so $( r_+,s_+)\rightarrow 1$ uniformly.
\end{proof}
Thus, we define
\begin{equation}  \label{eq:projtr}
    \langle P_{z_2}^2\rangle :=1
\end{equation}
\begin{equation}  \label{eq:projn}
    \langle P_{z_2}^2P_{z_3}\dots P_{z_n}\rangle :=
    \langle P_{z_2}P_{z_3}\dots P_{z_n}\rangle.
\end{equation}
Applying the relation 7 given in Definition~\ref{def:algebra} to
(\ref{eq:projn}), we get the relation
\begin{equation}   \label{eq:proj}
    P_z^2=P_z,\ z\in S^2_{\infty}
\end{equation}
so (\ref{eq:projtr}) and (\ref{eq:proj}) allow us to extend the 
definition of the $n$-point function to the 1-point function
\begin{equation}  \label{eq:onep}
    \langle P_z\rangle:=1,
\end{equation}
and from this it follows that
\begin{equation}  \label{eq:onepder}
    \langle[\partial_z,P_z]\rangle=0=\langle[\partial_{\bar{z}},P_z]\rangle.
\end{equation}

As described in the introduction, expectation values are used to
calculate the constant $c$ in relation 8.  When $P_{z_1}P_{z_2}=0$,
the expectation values of both sides of 8 are zero, so we instead
choose $z_0$ so that $\langle P_{z_0}P_{z_1}bP_{z_2}\rangle\neq 0$. 
(By relation 7, $z_0$ always exists.)  Then
\begin{equation}
\langle P_{z_0}P_{z_1}aP_{z_2}\rangle=c(z_1,z_2,a,b)\langle
P_{z_0}P_{z_1}bP_{z_2}\rangle
\end{equation}
enables us to calculate $c(z_1,z_2,a,b)$.  This introduces the issue 
of consistency of the algebra since the constant $c(z_1,z_2,a,b)$ can be 
calculated in different ways.  The following lemma gives the required 
property of the $n$-point function.
\begin{lemma}
     For $a\in\A$, $\langle P_{z_0}P_{z_2}\rangle \langle
     P_{z_0}P_{z_1}P_{z_2}a\rangle=\langle
     P_{z_0}P_{z_1}P_{z_2}\rangle \langle P_{z_2}aP_{z_0}\rangle.$
\end{lemma}
\begin{proof}
    For $a=P_{z_3}P_{z_4}\dots P_{z_n}$, where $z_i\neq z_{i+1}$, this
    follows simply from the definition.  Taking limits and derivatives 
    gives the result for general $a\in\A$.
\end{proof}

The Bogomolny equation implies that the Higgs field $\Phi$ satisfies a
maximum principle $\|\Phi\|<m$ where $m$ is the mass of the monopole.
This leads to a type of dissipative behaviour of
$(\partial_t^A-i\Phi)$ which can be used to show:
\begin{lemma}  \label{th:ineq}
For $z_i\neq z_{i+1}$, $z_n\neq z_1$, $|\langle P_{z_1}\dots
P_{z_n}\rangle|< 1.$
\end{lemma}
\begin{proof}
Since $\langle P_{z_1}\dots P_{z_n}\rangle=( r_{12},s_{12})(
r_{23}, s_{23})\dots( r_{n1},s_{n1})$ it is
sufficient to show along any geodesic that the solutions $s_+,r_+$ of
(\ref{eq:scat}) satisfy $|( r_+,s_+)|<1$, and in fact
\begin{eqnarray*}
|( r_+,s_+)|^2&=&\lim_{t\rightarrow-\infty}|( r_+(t),s_+(t)
)|^2\\
&=&\lim_{t\rightarrow-\infty}|(\exp(-mt)r_+(t),\exp(mt)s_+(t)
)|^2\\
&\leq&\lim_{t\rightarrow-\infty}\|\exp(-mt)r_+(t)\|^2\|\exp(mt)s_+(t)\|^2\\
&=&\lim_{t\rightarrow-\infty}\|\exp(mt)s_+(t)\|^2
\end{eqnarray*}
so we will show that $\lim_{t\rightarrow-\infty}\|\exp(mt)s_+(t)\|^2<1$.
We have 
\begin{eqnarray*}
|\partial_t\|s_+\|^2|&=&|((\partial_t^A+i\Phi)s,
s)+( s,(\partial_t^A-i\Phi)s)|\\
&=& |(2i\Phi s,s)|< 2m\|s,s\|^2
\end{eqnarray*}
where the last inequality uses the maximum principle $|\Phi|<m$.
Thus
\[\partial_t\|\exp(mt)s_+\|^2=(2m\|s,s\|^2+\partial_t\|s_+\|^2)\exp(2mt)
> 0.\] 
So the function $\|\exp(mt)s_+\|^2$ is strictly increasing, and by
construction of $s_+$,
$\lim_{t\rightarrow\infty}\|\exp(mt)s_+(t)\|^2=1$ yielding the
required inequality
\[\lim_{t\rightarrow-\infty}\|\exp(mt)s_+(t)\|^2<1.\]
\end{proof}

\begin{cor} \label{th:emb}
$P_{z_1}\neq P_{z_2}$ for $z_1\neq z_2$.
\end{cor}   
\begin{proof}
If $P_{z_1}=P_{z_2}$ then $\langle P_{z_1}P_{z_2}\rangle=\langle
P_{z_2}^2\rangle=1$ which contradicts Lemma~\ref{th:ineq}.
\end{proof}

Until now, we have only used the fact that $(A,\Phi)$ satisfies the
Bogomolny equation very mildly via the maximum principle for $\Phi$
and Rade's estimates for the monopole field.  Using the full structure
of the Bogomolny equation we can show that the assignment $z\mapsto
P_z$ possesses a holomorphic property.  It is used to prove the most
striking properties of the 2-point function and the existence of a
useful finite dimensional representation of $\A$.

With respect to particular local coordinate systems, the Bogomolny
equation $d_A\Phi=*F_A$ decomposes into a holomorphic part and a
``moment map'' part.  Specifically, this occurs for local coordinate
systems that reflect the holomorphic structure on the variety of
geodesics.  Two examples of this are the local coordinates $(t,z)$ in
$\bbH^3$ obtained from a family of geodesics, each parametrised by
$t$, traveling from the fixed $w\in S^2_{\infty}$ to the varying
$z\in S^2_{\infty}$, and the local coordinates $(t,w)$ in $\bbH^3$
obtained from a family of geodesics, each parametrised by $t$,
traveling from the varying $w\in S^2_{\infty}$ to the fixed $z\in
S^2_{\infty}$.  The Bogomolny equation decomposes into
$[\partial_{\bar{z}}^A,\partial_t^A-i\Phi]=0$, or equivalently,
$[\partial_z^A,\partial_t^A+i\Phi]=0$, and a second equation which we
will omit.  Similarly, $[\partial_w^A,\partial_t^A-i\Phi]=0$ and the
equivalent $[\partial_{\bar{w}}^A,\partial_t^A+i\Phi]=0$ are
consequences of the Bogomolny equation.  In particular, if $r_+$ and
$s_+$ are decaying solutions of (\ref{eq:scat}) then
\begin{equation}  \label{eq:covder}
    \begin{array}{cc}\partial_z^Ar_+=\mu_1(w,z)r_+,& \partial_{\bar{z}}^As_+
    =\lambda_1(w,z)s_+\\
    \partial_{\bar{w}}^Ar_+=\mu_2(w,z)r_+,&
    \partial_w^As_+=\lambda_2(w,z)s_+. 
    \end{array}
\end{equation}    
for (scalar) coefficients $\mu_i$, $\lambda_i$ {\em independent} of $t$. 
(These are used to obtain a
holomorphic bundle, with sub-line bundles on the variety of geodesics
of $\bbH^3$, \cite{AtiMag,HitCon}.)

\begin{prop}  \label{th:hol}
$P_z[\partial_{\bar{z}},P_z]=[\partial_{\bar{z}},P_z].$
\end{prop}
\begin{proof}
Consider the 3-point function
\begin{equation}  \label{eq:vary} 
\langle P_{z_1}P_{z_2}P_{z_3}\rangle=\langle 
P_{z_1}P_{z_2}P_z\rangle=( r_{12},s_{12})
( r_{23}(z),s_{23}(z))( r_{31}(z),s_{31}(z))
\end{equation}
where $z_3=z$ is allowed to vary, $z_1$ and $z_2$ are fixed and
different from $z$, and $r_{ij}$, $s_{ij}$ are the solutions of
(\ref{eq:scat}) along the geodesic running from $z_i$ to $z_j$.  We
have
\begin{equation}  \label{eq:der}
\partial_{\bar{z}}\langle P_{z_1}P_{z_2}P_z\rangle=\langle
[\partial_{\bar{z}},P_{z_1}P_{z_2}P_z]\rangle=\langle 
P_{z_1}P_{z_2}[\partial_{\bar{z}},P_z]\rangle
\end{equation}
and this will be used to characterise $P_z[\partial_{\bar{z}},P_z]$. 

By (\ref{eq:covder}) the Bogomolny equation implies that
$\partial_z^Ar_{23}(z)=\mu(z)r_{23}(z)$ with $z$ dependent
coefficient, and $\partial_{\bar{z}}^As_{23}(z)=\lambda(z)s_{23}(z)$,
since we are moving only one end of the geodesic.  The limit
$\lim_{t\rightarrow-\infty}r_{23}(z)$ is independent of $z$ so we can
multiply $r_{23}(z)$ by a function depending on $z$ and arrange that
$\mu(z)=0$, whilst preserving its normalisation at $t=-\infty$.  (We
cannot do the same for $\lambda(z)$.)  Thus
\begin{eqnarray*}
\partial_{\bar{z}}( r_{23}(z),s_{23}(z)) &=& 
(\partial_z^Ar_{23}(z),s_{23}(z))+ 
( r_{23}(z),\partial_{\bar{z}}^As_{23}(z)) \\
&=&\lambda(z)( r_{23}(z),s_{23}(z)).
\end{eqnarray*}
If we differentiate (\ref{eq:vary}) then we get
\begin{eqnarray*}
\langle P_{z_1}P_{z_2}[\partial_{\bar{z}},P_z]\rangle&=&\lambda(z)(
r_{12},s_{12})(r_{23}(z),s_{23}(z))(r_{31}(z),s_{31}(z))\\&&\ +(
r_{12},s_{12})(r_{23}(z),s_{23}(z))\partial_{\bar{z}}(r_{31}(z),s_{31}(z)).
\end{eqnarray*}
Let $z_2\rightarrow z$.  Then as shown in the proof of
Lemma~\ref{th:square}, $(r_{23},s_{23})\rightarrow 1$ so
\begin{eqnarray*}
\lim_{z_2\rightarrow z}\langle P_{z_1}
P_{z_2}[\partial_{\bar{z}},P_z]\rangle&=&\lambda(z)(
r_{12},s_{12})(r_{31},s_{31})+(
r_{12},s_{12})\partial_{\bar{z}}(r_{31},s_{31})\\
&=&\langle P_{z_1}[\partial_{\bar{z}},P_z]\rangle.
\end{eqnarray*}
Hence $\langle P_{z_1}P_z[\partial_{\bar{z}},P_z]\rangle=\langle
P_{z_1}[\partial_{\bar{z}},P_z]\rangle$ for
all $z_1$, so we get the relation \[
P_z[\partial_{\bar{z}},P_z]=[\partial_{\bar{z}},P_z]\] as required.
\end{proof}
The relation is equivalent to 
\[[\partial_{\bar{z}},P_z]P_z=0=P_z[\partial_z,P_z].\] 
We call this a holomorphic relation since it gives a type of
integrability condition whereby $\bar{\partial}$ is preserved by $P$,
and since it will translate precisely to an integrability condition
when we construct a representation of $\A$.  

By Proposition~\ref{th:hol} $\partial_{\bar{z}}\langle
P_wP_z\rangle=\langle P_wP_z [\partial_{\bar{z}},P_z]\rangle$ so that
\begin{equation*}
\langle P_wP_z\rangle=0\ \Rightarrow\ \partial_{\bar{z}}\langle 
P_wP_z\rangle=0.
\end{equation*}
This suggests that it might be fruitful to take some type of log
derivative of the 2-point function.  In the remainder of this section
we will show that the 2-point function, when viewed appropriately, is
both a defining function for the spectral curve of the monopole and a
Hermitian metric for the connection on the conformal boundary two-sphere.

\begin{lemma}  \label{th:proplam}
    The function 
    \begin{equation}  \label{eq:lambda}
	\lambda(w,z)=(1/2)\partial_{\bar{z}}\ln\langle P_wP_z\rangle
    \end{equation}
    satisfies
    (i) $\lambda(z,z)=0$, (ii) $\lambda(w,z)$ is meromorphic in $w$,
    and (iii) $\partial_z\lambda(w,z)$ is real and independent of $w$.
\end{lemma}
\begin{proof}
    (i) We have $2\lambda(z,z)=\lim_{w\rightarrow
    z}\partial_{\bar{z}}\ln\langle P_wP_z\rangle=\lim_{w\rightarrow
    z}\langle P_w[\partial_{\bar{z}},P_z]\rangle/\langle
    P_wP_z\rangle$.  This can be simplified to $\langle
    P_z[\partial_{\bar{z}},P_z]\rangle=\langle
    [\partial_{\bar{z}},P_z]\rangle=0$ by Proposition~\ref{th:hol} and
    (\ref{eq:onepder}).

    (ii) In an open set of $\bbC\bbP^1\times\bbC\bbP^1$ choose
    solutions of (\ref{eq:scat}) normalised by (\ref{eq:normal}) so
    that $\lim_{t\rightarrow\infty}\exp(mt)s_+(w,z)$ is independent of
    $w$ and $\lim_{t\rightarrow-\infty}\exp(-mt)r_+(w,z)$ is
    independent of $z$.  (To achieve this choose a normalised solution
    of (\ref{eq:scat}) $s_+(w_0,z)$ for a fixed $w=w_0$ and use
    $\lim_{t\rightarrow\infty}\exp(mt)s_+(w_0,z)=
    \lim_{t\rightarrow\infty}\exp(mt)s_+(w,z)$ to define $s_+(w,z)$
    for nearby $w$.  Do the same for $r_+(w,z)$ around $z_0$.)
    Therefore, $\partial_w^As_+=0=\partial_z^Ar_+$ and
    $\partial_{\bar{z}}^As_+=\mu_1(z)s_+$,
    $\partial_{\bar{w}}^Ar_+=\mu_2(w)r_+$ for $\mu_1(z)$ independent
    of $w$ and $\mu_2(w)$ independent of $z$, since we can calculate
    the coefficients in (\ref{eq:covder}) in the infinite limits.
    Then $\lambda(w,z)$ is holomorphic away from it singular points:
    \begin{eqnarray*} 
        \partial_{\bar{w}}\partial_{\bar{z}}\ln\langle
        P_wP_z\rangle&=&\partial_{\bar{w}}\partial_{\bar{z}}
        \ln|(r_+,s_+)|^2=\partial_{\bar{w}}\partial_{\bar{z}}
	(\ln(r_+,s_+)+\ln(s_+,r_+))\\&=& \partial_{\bar{w}}\mu_1(z)+
        \partial_{\bar{z}}\mu_2(w)=0.  
    \end{eqnarray*} 
    The meromorphicity of $\lambda(w,z)$ at the singular points
    follows from Lemma~\ref{th:van} where it is shown that $\langle
    P_wP_z\rangle$ vanishes like $|\psi(w,z)|^2$ for a locally defined
    function $\psi(w,z)$, anti-holomorphic in $w$ and holomorphic in
    $z$, with the same zero set as $\langle P_wP_z\rangle$.  Thus,
    ignoring the finite parts of $\lambda(w,z)$ its singular part
    looks like $\partial_{\bar{z}}\ln\overline{\psi(w,z)}=
    \partial_{\bar{z}}\overline{\psi(w,z)}/\overline{\psi(w,z)}$ 
    which has a pole in the variable $w$.

    (iii) For $\lambda(w,z)$ defined in (\ref{eq:lambda}) we can
    choose a local gauge in which
    \[\partial_{\bar{z}}^As_+(w,z)=\lambda(w,z)s_+(w,z)\] as follows. 
    Choose $r_+(w,z)$ so that $\partial_z^Ar_+(w,z)=0$ (as in (ii).) 
    Now choose $s_+(w,z)$ so that $(r_+(w,z),s_+(w,z))$ is real.  This
    uniquely determines $s_+$ up to a constant $U(1)$ gauge
    transformation given by the ambiguity in the phase of $r_+$.  Then
    \begin{eqnarray*}
	\partial_{\bar{z}}\langle P_wP_z\rangle&=&\partial_{\bar{z}}|(
	r_+,s_+)|^2=\partial_{\bar{z}}( r_+,s_+)^2\\
	&=&2(r_+,s_+)(r_+,\partial^A_{\bar{z}}s_+)\\
	&=&2\lambda(w,z)(r_+,s_+)^2=2\lambda(w,z)\langle P_wP_z\rangle.
    \end{eqnarray*}
    For $w'\neq w$ choose the solutions of (\ref{eq:scat}) normalised
    by (\ref{eq:normal}) along each family of geodesics, respectively
    $r'_+(z)$, $s'_+(z)$, $r_+(z)$ and $s_+(z)$, so that
    $\partial_z^Ar_+(z)=0$ and $(r_+(z),s_+(z))\in\bbR$, and
    $\partial_z^Ar'_+(z)=0$ and $\partial_{\bar{z}}^As'_+(z)=
    \lambda(w,z)s'_+(z)$ (define $s'_+$ via
    $\lim_{t\rightarrow\infty}\exp(mt)s'_+(z,t)=\lim_{t\rightarrow\infty}
    \exp(mt)s_+(z,t)$.)  We can compare $\lambda(w',z)$ and
    $\lambda(w,z)$ by defining $\theta(z)$ so that
    $(r'_+(z),\exp(i\theta(z))s'_+(z))\in\bbR$, then
    \[\lambda(w',z)=\lambda(w,z)+i\partial_{\bar{z}}\theta(z).\] In
    particular the expression in (iii) is independent of $w$:
    \begin{eqnarray*}
	\partial_z\partial_{\bar{z}}\ln \langle P_{w'}P_z\rangle
	dzd\bar{z}
	&=&2(\partial_z\lambda(w',z)+\partial_{\bar{z}}\bar{\lambda}(w',z))
	dzd\bar{z}\\
	&=&2(\partial_z\lambda(w,z)+\partial_{\bar{z}}\bar{\lambda}(w,z)+
	i\partial_{\bar{z}}\partial_z\theta-i\partial_z\partial_{\bar{z}}
	\theta)dzd\bar{z}\\
	&=&\partial_z\partial_{\bar{z}}\ln \langle P_wP_z\rangle
	dzd\bar{z}
    \end{eqnarray*}
    and real since it is the Laplacian of a real-valued function.
\end{proof}

If we replace $w$ in $\lambda(w,z)$ by its antipodal point
$\hat{w}=-1/{\bar{w}}$ then although $\lambda(\hat{w},z)$ is defined
only outside the set $\langle P_{\hat{w}}P_z\rangle=0$, the 2-form
\begin{eqnarray*} 
    (1/2)\partial\bar{\partial}\ln\langle
    P_{\hat{w}}P_z\rangle&=& \partial_z\lambda(\hat{w},z)dz
    d\bar{z}+\partial_w\bar{\lambda} (z,\hat{w})dwd\hat{w}\\&&\ \ \ \
    +\partial_w\lambda (\hat{w},z)dw
    d\bar{z}+\partial_z\bar{\lambda}(z,\hat{w})dzd \hat{w}
\end{eqnarray*} 
is well-defined everywhere.  To see this, first notice that the term
$\partial_w\lambda (\hat{w},z)$ vanishes by Lemma~\ref{th:proplam}
(ii) and for the same reason $\partial_z\bar{\lambda}(z,\hat{w})dzd
\hat{w}$ vanishes.  The term $\partial_z\lambda(\hat{w},z)$ is
independent of $w$ by Lemma~\ref{th:proplam} (iii) so in particular it
is well-defined everywhere since for any $z$ we can choose a $w$ such
that $\langle P_wP_z\rangle\neq 0$, and the same is true of
$\partial_w\bar{\lambda} (z,\hat{w})dwd\hat{w}$.  Thus the 2-form
$\partial\bar{\partial}\ln\langle P_{\hat{w}}P_z\rangle$ is a
well-defined closed $(1,1)$ form.  We can use this to prove that the
zero set of the real-valued function $\langle P_{\hat{w}}P_z\rangle$
is holomorphic, but instead we will rely on known facts about the 
spectral curve of the monopole.

\begin{prop}   \label{th:spec}
    The spectral curve of the monopole is encoded in the 2-point
    function.  It is given by \[\Sigma=\{
    (w,z)\in\bbC\bbP^1\times\bbC\bbP^1\ |\ \langle
    P_{\hat{w}}P_z\rangle =0\}\] for $\hat{w}$ the antipodal point of
    $w$ in $\bbC\bbP^1$.
\end{prop}
\begin{proof}
    This follows from the simple fact that $\langle
    P_{\hat{w}}P_z\rangle=0$ precisely when the solutions $r_+,s_+$ of
    (\ref{eq:scat}) decay at both ends, which is the same condition
    for a geodesic to lie in the spectral curve.  Notice that the
    invariance of $\Sigma$ under the real structure
    $(w,z)\mapsto(\hat{z},\hat{w})$ extends to the 2-point function
    since $\langle P_{\hat{w}}P_z\rangle=\langle
    P_zP_{\hat{w}}\rangle$.
\end{proof}
We could have equivalently stated Proposition~\ref{th:spec} in terms
of the multiplication operation of the algebra $\cS(A,\Phi)$ in place
of the 2-point function since $P_{\hat{w}}P_z=0$ is equivalent to
$\langle P_{\hat{w}}P_z\rangle=0$.

\begin{prop}  \label{th:conenc}
    The connection on the conformal boundary two-sphere is encoded in
    the 2-point function by $\lambda(w,z)
    =(1/2)\partial_{\bar{z}}\ln\langle P_wP_z\rangle$ and \[
    A_{\infty}=\lambda(w,z)d\bar{z}-\bar{\lambda}(w,z)dz\] where $w$
    is fixed and gives a choice of gauge.  The curvature on the
    conformal boundary two-sphere is given by
    $F_{A_{\infty}}=-\langle[\partial_z,P_z][\partial_{\bar{z}},P_z]\rangle
    dzd\bar{z}$.
\end{prop}
\begin{proof}
    Fix $w$ and vary $z$.  The Bogomolny equation implies that the
    solution $s_+$ of (\ref{eq:scat}) normalised by (\ref{eq:normal})
    also satisfies $\partial_{\bar{z}}^As_+(z,t)=\lambda(z)s_+(z,t)$
    for $\lambda(z)$ independent of $t$.  In the limit, the section
    $\lim_{t\rightarrow\infty}\exp(mt)s_+(z,t)$ gives a unitary gauge
    for the connection on the conformal boundary two-sphere, and hence
    $\lambda(z)d\bar{z}$ is the $d\bar{z}$ component of $A_{\infty}$. 
    Any other choice of $s_+(z,t)$ satisfying (\ref{eq:normal})
    differs by $\exp(i\theta(z))$ and hence
    \[\lambda(z)\mapsto\lambda(z)+i\partial_{\bar{z}}\theta(z)\] which
    is a change of the $U(1)$ gauge.  In fact, without the
    normalisation (\ref{eq:normal}), the $\lambda(z)$ that arises
    gives the connection on the conformal boundary two-sphere which is
    Hermitian with respect to a Hermitian metric defined by
    $\lim_{t\rightarrow\infty}\|\exp(mt)s_+(z,t)\|^2$.  As in the
    proof of Lemma~\ref{th:proplam} we can choose $r_+(z)$ and
    $s_+(z)$ so that $\partial_z^Ar_+(z)=0$ and $(r_+(z),s_+(z))$ is
    real.  Then $\partial^A_{\bar{z}}s_+=\lambda(w,z)s_+$ so
    $\lambda(w,z)d\bar{z}$ gives the $(0,1)$ part of $A_{\infty}$ with
    respect to a well-defined $U(1)$ gauge (up to a constant gauge
    transformation) determined by the choice of $w$.  Thus the first
    part of the proposition is proven.

    The curvature is given by \[ F_{A_{\infty}} =
    (\partial_z\lambda(w,z)+\partial_{\bar{z}}
    \bar{\lambda}(w,z))dzd\bar{z}=\partial_z\partial_{\bar{z}}\ln
    \langle P_wP_z\rangle dzd\bar{z}\] since $\partial_z\lambda(w,z)$
    is real-valued, and
    \begin{eqnarray*}
	\partial_z\partial_{\bar{z}}\ln \langle
	P_wP_z\rangle&=&(\partial_z\partial_{\bar{z}}\langle
	P_wP_z\rangle)/\langle
	P_wP_z\rangle-(\partial_{\bar{z}}\langle
	P_wP_z\rangle\partial_z\langle P_wP_z\rangle)/\langle
	P_wP_z\rangle^2\\ &=& \langle
	P_w[\partial_z,[\partial_{\bar{z}},P_z]]\rangle)/\langle
	P_wP_z\rangle -\langle P_w[\partial_{\bar{z}},P_z]\rangle
	\langle P_w[\partial_z,P_z]\rangle/\langle P_wP_z\rangle^2.
    \end{eqnarray*}
    This is independent of $w$, since it is a gauge invariant 2-form
    or we see it explicitly in Lemma~\ref{th:proplam}.  Thus we can
    take the limit $w\rightarrow z$ and since $P_z[\partial_z,P_z]=0$
    the second term disappears to leave
    \[\partial_z\partial_{\bar{z}}\ln \langle P_wP_z\rangle= \langle
    P_z[\partial_z,[\partial_{\bar{z}},P_z]]\rangle.\] Since
    $0=\langle [\partial_{\bar{z}},P_z]]\rangle=\langle P_z
    [\partial_{\bar{z}},P_z]]\rangle$ then \[ 0=\partial_z\langle P_z
    [\partial_{\bar{z}},P_z]]\rangle=
    \langle[\partial_z,P_z][\partial_{\bar{z}},P_z]\rangle+\langle
    P_z[\partial_z,[\partial_{\bar{z}},P_z]]\rangle\] thus
    \[F_{A_{\infty}}=\partial_z\partial_{\bar{z}}\ln \langle
    P_wP_z\rangle
    dzd\bar{z}=-\langle[\partial_z,P_z][\partial_{\bar{z}},P_z]\rangle
    dzd\bar{z}.\]
\end{proof}
The construction of the gauge in which
$A_{\infty}=\lambda(w,z)d\bar{z}-\bar{\lambda}(w,z)dz$ breaks down if
$\langle P_wP_z\rangle=0$.  In that case, once $r_+(z)$ is chosen,
there is not a unique choice of $s_+(z)$ that satisfies $(
r_+(z),s_+(z))$ is real.  This simply says that the $U(1)$ gauge
defined by $w$ is well-defined, up to locally constant gauge
transformations, on the complement of the finite set of points $\{
z_1,\dots,z_k\}$ determined by $\langle P_wP_{z_i}\rangle=0$, or in
other words, $w$ defines a flat structure on a line bundle over
$S^2-\{ z_1,\dots,z_k\}$.

An understanding of the behaviour of $A_{\infty}$ with respect to the
gauge in Proposition~\ref{th:conenc} near its singularities is a key
ingredient in the proof that the connection on the conformal boundary
two-sphere uniquely determines the 2-point function.  Equivalently, we
must understand the behaviour of the 2-point function near its zero
set.

\begin{lemma}  \label{th:van}
    Near a point $(w_0,z_0)$ in the zero set, $\langle
    P_{\hat{w_0}}P_{z_0}\rangle=0$, the function $\langle
    P_{\hat{w}}P_z\rangle$ vanishes like $|\psi(w,z)|^2$, where
    $\psi(w,z)$ is a local holomorphic defining function for the zero
    set.
\end{lemma}
\begin{proof}
    In order to study the vanishing at $\langle
    P_{\hat{w_0}}P_{z_0}\rangle$, we may ignore the normalisation
    condition (\ref{eq:normal}) of solutions $r_+(\hat{w},z)$ and
    $s_+(\hat{w},z)$ of (\ref{eq:scat}) since that simply involves
    multiplying the solutions by non-vanishing functions.  Thus we may
    choose the solutions so that
    $\partial_z^Ar_+=0=\partial_{\bar{z}}^As_+$ and
    $\partial_{\bar{w}}^Ar_+=0=\partial_w^As_+$.  The inner product
    $(r_+(\hat{w},z),s_+(\hat{w},z))$ is generically a transverse
    local section of the line bundle $\mathcal{O}(k,k)$ so
    $|(r_+(\hat{w},z),s_+(\hat{w},z))|^2$ vanishes like
    $|\psi(w,z)|^2$ and so too does $\langle P_{\hat{w}}P_z\rangle$.
\end{proof}

We will summarise the properties of the gauge for $A_{\infty}$ in the
following proposition.
\begin{prop} \label{th:gauge}
    The $(0,1)$ part of $A_{\infty}$, given by
    $\eta_w(z)=\lambda(w,z)d\bar{z}$, satisfies the properties:
    \begin{enumerate}
	\item $\eta_w(z)$ is well-defined outside a set of points $\{
	z_1,\dots,z_k\}$; 
	\item $\eta_w(z)\sim\ln|z-z_i|^2d\bar{z}$ at each $z_i$; 
	\item $d\eta_w(z)$ is an imaginary valued 2-form;
	\item $\eta_w|_{z=z_0}$ is meromorphic in $w$; 
	\item $\eta_w(w)=0$.
    \end{enumerate}
    Furthermore, this $U(1)$ gauge is the unique gauge (up to a
    constant gauge transformation) satisfying properties 1-3.
\end{prop}
\begin{proof}
    The points $\{ z_1,\dots,z_k\}$ are determined by $\langle
    P_wP_{z_i}\rangle=0$ and Lemma~\ref{th:van} determines the
    behaviour of the singularities there.  Properties 3, 4 and 5
    follow from Lemma~\ref{th:proplam}.  Any other 1-form with these
    properties must differ from $\eta(z)$ by
    $i\partial_{\bar{z}}\theta(z)d\bar{z}$ for a real-valued function
    $\theta(z)$.  By 1, $\theta(z)$ is a function defined outside the
    set of points $\{ z_1,\dots,z_k\}$ and by 2 and 3 it is bounded
    and harmonic and hence constant.  Thus
    $i\partial_{\bar{z}}\theta(z)d\bar{z}=0$ and the properties
    uniquely determine $\eta$.
\end{proof}

Properties 4 and 5 are automatically satisfied by any $\eta(z)$
satisfying 1, 2 and 3.  This suggests that the connection on the
conformal boundary two-sphere in some sense feels the spectral curve. 
The next proposition will prove that the connection on the conformal
boundary two-sphere does determine the 2-point function and hence the
spectral curve.

\begin{prop}  \label{th:infalg}
    The connection on the conformal boundary two-sphere uniquely
    determines the 2-point function.
\end{prop}
\begin{proof}
    Suppose we have two monopoles $(A,\Phi)$ and $(A',\Phi')$ with
    respective algebras consisting of elements $P_z$ and $P'_z$.  Fix
    $w$ and vary $z$.  The two monopoles have the same connection on
    the conformal boundary two-sphere precisely when
    \begin{equation}  \label{eq:difp}
	\ln\langle P_wP_z\rangle-\ln\langle P'_wP'_z\rangle
    \end{equation}
    is harmonic in $z,\bar{z}$, since the curvatures of the
    connections on the conformal boundary two-sphere must coincide.

    With respect to a local trivialisation of ${\mathcal O}(k,k)$ in
    the neighbourhood of a point on $\bar{\Delta}$ denote by
    $\Psi(w,z)$ a section with zero set the spectral curve of
    $(A,\Phi)$, and similarly $\Psi'(w,z)$ for $(A',\Phi')$.  (We are
    using the fact that a section of ${\mathcal O}(k,k)$ over
    $\bbC\bbP^1\times\bbC\bbP^1-\bar{\Delta}$ extends to a section
    over $\bbC\bbP^1\times\bbC\bbP^1$.  This follows from considering
    the $k$th formal neighbourhood of the diagonal 
    $\Delta\subset\bbC\bbP^1\times\bbC\bbP^1-\bar{\Delta}$.)  Then
    \begin{equation} \label{eq:canlog} \ln\langle
    P_{\hat{w}}P_z\rangle-\ln\langle P'_{\hat{w}}P'_z\rangle
    +\ln|\Psi'(w,z)|^2/|\Psi(w,z)|^2
    =\ln|\Psi'(w,\hat{w})|^2/|\Psi(w,\hat{w})|^2 \end{equation} since
    the left hand side of (\ref{eq:canlog}) is well-defined
    everywhere, i.e. we have canceled singularities, and for fixed $w$
    it is harmonic in $z,\bar{z}$.  Hence it is constant in $z$ and
    when we evaluate at $z=\hat{w}$ we get the right hand side.

    Now fix $z$ and take $\partial_w\partial_{\bar{w}}$ of both sides
    of (\ref{eq:canlog}).  The left hand side vanishes since
    (\ref{eq:difp}) is also harmonic in $w,\bar{w}$ by symmetry.  Thus
    $\ln|\Psi(w,\hat{w})|^2- \ln|\Psi'(w,\hat{w})|^2$ is harmonic in
    $w,\bar{w}$.  If $\xi(w)$ is harmonic then it is the sum of a
    holomorphic and anti-holomorphic function since $\xi+i\rho$ is
    holomorphic for some (locally defined $\rho(w)$) and $\xi-i\rho$
    is anti-holomorphic.  We can choose $\Psi$ to be real and positive
    on $\bar{\Delta}$ so $\ln|\Psi(w,\hat{w})|^2=2\ln\Psi(w,\hat{w})$
    and similarly for $\Psi'$.  Thus \[
    \Psi(w,\hat{w})=g_1(w)g_2(\hat{w})\Psi'(w,\hat{w})\] for $g_1(w)$
    holomorphic and $g_2(\hat{w})$ anti-holomorphic.  We conclude that
    \[\Psi(w,z)=g_1(w)g_2(z)\Psi'(w,z)\] since the real analytic
    function $\Psi(w,\hat{w})$ on $\bar{\Delta}$ has a unique
    extension in a neighbourhood of
    $\bar{\Delta}\subset\bbC\bbP^1\times\bbC\bbP^1$.  But then $g_1$
    and $g_2$ are both constant since $\Psi_{|\bar{\Delta}}\neq 0$ so
    the zero set of $\Psi$ cannot contain lines $w=w_0$ or $z=z_0$.

    Thus, $\langle P_wP_z\rangle-\langle P'_wP'_z\rangle$ is constant
    and hence $0$ since they agree on $w=z$.
\end{proof}
{\em Remark.} This completes the proof of Theorem~\ref{th:2ptinf} and
Corollary~\ref{th:conj}.  On closer observation, one soon realises
that one of the key facts in the proof of
Proposition~\ref{th:infalg}---$\Psi(w,\hat{w})$, defined up to
multiplication by the norm squared of a holomorphic function, uniquely
determines $\Psi(w,z)$ up to a constant---leads to another proof
of Corollary~\ref{th:conj}.  This viewpoint is taken in \cite{MNoHyp}.

\section{Representation}
Consider a representation of $\cS(A,\Phi)$ on a Hilbert space $H$ that
satisfies
\begin{equation}  \label{eq:rep}
    \langle a\rangle=tr\ a\ {\rm and}\ a^*\ {\rm is\ the\ adjoint\ 
    of\ } a,
\end{equation}
where we abuse notation and denote $a\in\A$ to also mean its image in
the space of endomorphisms of $H$.  The properties $P_z^2=P_z=P_z^*$
and $tr\ P_z=\langle P_z\rangle=1$ imply that $P_z$ is a {\em
projection} with one-dimensional image.  The image of each projection
is a line in $H$ so each $P_z$ corresponds to a point in $\bbP H$ and
we have a map $q:S^2_{\infty}\rightarrow\bbP H$ defined by $q(z)={\rm
im}\ P(z)$.  In this section we will describe the properties of $\A$
in terms of the map $q$.  We will defer the proof of existence of a
representation until the end of the section.  Let $k$ be the charge of
the monopole.
\begin{prop}   \label{th:rephol}
    A representation of $\cS(A,\Phi)$ on a Hilbert space $H$
    satisfying (\ref{eq:rep}) gives rise to a 1-1 degree $k$
    holomorphic map $q:S^2_{\infty}\rightarrow\bbC\bbP^k$.
\end{prop}
\begin{proof}
    We will use $|q(z)\rangle$ to label a unit vector in the line
    $q(z)={\rm im}\ P(z)\subset H$, and $\langle q(z)|$ its conjugate
    transpose, so $\langle q(z)|q(z)\rangle=1$.  Thus $|q(z)\rangle$
    is still ambiguous up to a phase, although \[ |q(z)\rangle\langle
    q(z)|=P_z\] is well-defined.
    
    To show that $q(z)$ is smooth at $z_0$, choose a $w$ so that
    $P_wP_{z_0}\neq 0$ and choose a neighbourhood $U$ of $z_0$ so that
    $P_wP_z\neq 0$ for $z\in U$.  Then fix a unit vector
    $|q(w)\rangle$ and for each $z\in U$ choose a unit vector
    $|q(z)\rangle$ so that $\langle q(w)|q(z)\rangle$ is real.  Then
    by Lemma~\ref{th:cont} $\langle P_wP_z\rangle=tr\ P_wP_z=\langle
    q(w)|q(z)\rangle^2$ is smooth in $z$ so $\langle q(w)|q(z)\rangle$
    is smooth in $z$.  Thus the component $P_wq(z)$ of $q(z)$ is
    smooth.  This is true for almost all $w$ so $q(z)$ is smooth on
    the linear span of the image of $q$.  We may replace $H$ by this
    linear span, since the representation annihilates the complement. 
    Thus $q(z)$ is a smooth map.
    
    The holomorphicity of $q(z)$ is equivalent to the property
    $P_z[\partial_{\bar{z}},P_z]=[\partial_{\bar{z}},P_z]$ proven in
    Proposition~\ref{th:hol}.  This can be seen by setting
    $P_z=|q(z)\rangle\langle q(z)|$.  Then \[ |q(z)\rangle\langle
    q(z)|(|\partial_{\bar{z}}q(z)\rangle\langle q(z)|+
    |q(z)\rangle\langle\partial_zq(z)|)=(|\partial_{\bar{z}}q(z)\rangle\langle
    q(z)|+|q(z)\rangle\langle \partial_zq(z)|)\] \[ \Rightarrow\ \
    |q(z)\rangle\langle q(z)|\partial_{\bar{z}}q(z)\rangle \langle
    q(z)|= |\partial_{\bar{z}}q(z)\rangle\langle q(z)|\] and by acting
    on the left by any vector orthogonal to $|q(z)\rangle$ we see that
    \[\partial_{\bar{z}}|q(z)\rangle=\lambda(z)|q(z)\rangle\] for some
    function $\lambda(z)$, so $q(z)$ is holomorphic.  (We use
    $\partial_{\bar{z}}|q(z)\rangle$ and
    $|\partial_{\bar{z}}q(z)\rangle$ to mean the same thing.)
    
    The degree of $q(z)$ is obtained by intersecting its image with a
    hyperplane.  This corresponds to asking for the number of
    solutions $z$ to $P_wP_z=0$ for a generic $w$, which is $k$, the
    charge of the monopole.  Furthermore, the degree of $q(z)$
    determines an upper bound for the dimension of the span of its
    image, thus $q:S^2_{\infty}\rightarrow\bbC\bbP^k\subset\bbP H$.  The map
    $q(z)$ is one-to-one since the proof of Corollary~\ref{th:emb}
    shows not only that $P_w\neq P_z$ in $\A$ but also that their
    images under the representation are unequal via $tr\ P_wP_z<1$.
\end{proof}

\begin{prop}   \label{th:orth}
    The spectral curve of a charge $k$ $SU(2)$ hyperbolic monopole
    with associated holomorphic sphere $q:S^2_{\infty}\rightarrow\bbC\bbP^k$ is
    given by \[\Sigma=\{ (w,z)\in\bbC\bbP^1\times\bbC\bbP^1\ |\ (
    q(\hat{w}),q(z))=0\}\] where $\hat{w}$ is the antipodal point of
    $w$ and $(\cdot,\cdot)$ is the natural Hermitian product on
    $\bbC^{k+1}$.  Equivalently, $w^k( q(\hat{w}),q(z))=\psi(w,z)$,
    the defining polynomial of $\Sigma$.
\end{prop}
\begin{proof}
    This is simply a restatement of Proposition~\ref{th:spec} since
    the product of two projections is zero precisely when their images
    are orthogonal.  The function $( q(\hat{w}),q(z))$ is quite
    different from the corresponding function $\langle
    P_{\hat{w}}P_z\rangle$.  In particular it is holomorphic, and
    hence can be represented by a polynomial.
\end{proof}

Recall from \cite{ABrBou} that to an $SU(2)$ integral mass charge $k$
hyperbolic monopole one can associate a solution of the discrete Nahm
equations.  In the following $m\in\bbZ+1/2$.
\[\begin{array}{ll}
\gamma_j=\gamma_{-j}^T& -2m+2\leq j\leq 2m-2,\ j\ {\rm odd} \\
\beta_j=\beta_{-j}^T& -2m+1\leq j\leq 2m-1,\ j\ {\rm even} \\
\beta_{j-1}\gamma_j-\gamma_j\beta_{j+1}=0&-2m+2\leq j\leq 2m-2,\ j\ {\rm odd}\\
\lbrack\beta_j^*,\beta_j\rbrack+\gamma_{j-1}^*\gamma_{j-1}-\gamma_{j+1}
\gamma_{j+1}^*=0& -2m+3\leq j\leq 2m-3,\ j\ {\rm even} \\
\lbrack\beta_{2m-1},\beta_{2m-1}^*\rbrack
+v\bar{v}^T-\gamma_{2m-2}^*\gamma_{2m-2}=0&
\end{array}\]
where $\beta_i,\gamma_j\in{\bf gl}(k,\bbC)$ and $v\in\bbC^k$ admit an
action of $\{ g_j\in U(k)\ |\ j=-2m+1,-2m+3,\dots,0,\dots,2m-3,2m-1,\
g_j=\bar{g}_{-j}\}$ by
\begin{eqnarray*}
\beta_j&\mapsto&g_j\beta_j g_j^{-1}\\
\gamma_j&\mapsto&g_{j-1}\gamma_j g_{j+1}^{-1}\\
v&\mapsto&g_{2m-1}v
\end{eqnarray*}
(Note that we have replaced $v$ with $v^T$ from \cite{ABrBou} so that
the vector $v$ is a column vector and matrices act on its left.)  The
pair $(\beta_{-2m+1},v)$ determines the full solution of the discrete
Nahm equations.  It was shown in \cite{ABrBou} that the map
\begin{equation}  \label{eq:monad} 
\left(\begin{array}{c}\beta_{-2m+1}-z\\v^T\end{array}\right):
\bbC^k\rightarrow\bbC^{k+1}
\end{equation}
is a monad on $S^2$ which determines the boundary value of the
hyperbolic monopole.  The monad can be interpreted as a degree $k$
holomorphic map $\beta:S^2\rightarrow\bbC\bbP^k$ given explicitly by
\begin{equation}  \label{eq:monmap}
\beta(z)=\left(\begin{array}{c}-\det(\beta_{-2m+1}-z)\cdot
(\beta_{-2m+1}^T-z)^{-1}v\\\det(\beta_{-2m+1}-z)\end{array}\right).
\end{equation}
The map is well-defined up to the $U(k)$ action on the first $k$
coordinates, since $\beta_{-2m+1}$ admits a $U(k)$ action.  The map
$\beta$ has the properties that the pull-back of the K\"ahler form
$\beta^*\omega$ gives the curvature of the monopole on the conformal
boundary two-sphere (and hence its gauge equivalence class). 
Furthermore, by a theorem of Calabi the pull-back of the K\"ahler
form, and hence the curvature of the monopole on the conformal
boundary two-sphere, uniquely determines the map $\beta$.  Thus the
boundary value of the monopole determines the monopole.

\begin{prop}   \label{th:specdn}
The spectral curve of $(A,\Phi)$ is given by
\[\Sigma=\{ (w,z)\in\bbC\bbP^1\times\bbC\bbP^1\ |\ (
\beta(\hat{w}),\beta(z))=0\}.\]
\end{prop}
\begin{proof}
This is a simple result from linear algebra.
For any two vectors $u,v\in\bbC^n$,
\begin{equation}  \label{eq:linalg}
\det(1+u\bar{v}^T)=1+( v,u)
\end{equation}
since $(u,v)\mapsto(g^{-1}u,\bar{g}^Tv)$ preserves (\ref{eq:linalg})
for any $g\in GL(n,\bbC)$, so we may assume $u=(1,0,0,\dots)$, in
which case (\ref{eq:linalg}) is easy.

Put $d(w,z)=\det(\bar{\beta}_{-2m+1}+1/w)\det(\beta_{-2m+1}-z)$ for
ease in reading the next set of formulae.
\begin{eqnarray*}
(\beta(\hat{w}),\beta(z))&=&
% \det(\bar{\beta}_{-2m+1}+1/w)\det(\beta_{-2m+1}-z)
d(w,z)(\bar{v}^T(\bar{\beta}_{-2m+1}+1/w)^{-1}(\beta_{-2m+1}^T-z)v+1)\\
&=&
% \det(\bar{\beta}_{-2m+1}+1/w)\det(\beta_{-2m+1}-z)
d(w,z)\det(1+(\beta_{-2m+1}^T
-z)v\bar{v}^T(\bar{\beta}_{-2m+1}+1/w)^{-1})\ \ {\rm by}\ (\ref{eq:linalg})\\
&=&\det((\beta_{-2m+1}-z)(\bar{\beta}_{-2m+1}+1/w)+v\bar{v}^T)
\end{eqnarray*}
and the last expression defines the spectral curve by specialising the
expression in \cite{MSiCom} to the boundary value of the discrete Nahm
equations.
\end{proof}

\begin{cor}
    For half-integer mass, the holomorphic map
    $q:S^2_{\infty}\rightarrow\bbC\bbP^k$ associated to the algebra
    $\cS(A,\Phi)$ coincides up to the action of $U(k+1)$ on its image
    with the holomorphic map $\beta:S^2\rightarrow\bbC\bbP^k$ arising
    from the discrete Nahm equations.
\end{cor}
Strictly, we should say that in the $U(k+1)$ orbit of the map
$q:S^2_{\infty}\rightarrow\bbC\bbP^k$ associated to the algebra $\cS$, there is a
$U(k)$ orbit of the map $\beta$.
\begin{proof}
    The expressions \[w^k( \beta(\hat{w}),\beta(z))\ \ \ {\rm and}\ \
    \ w^k( q(\hat{w}),q(z))\] coincide since they both define
    holomorphic sections of $\cO(k,k)$ with the same zero set.  Thus
    $\beta(z)=uq(z)$ for some $u\in U(k+1)$.
\end{proof}
{\em Remark.}  Another corollary of Proposition~\ref{th:specdn} is a
new proof of the fact that the boundary value of the monopole
determines the monopole when the mass is a half integer.

\begin{prop}  \label{th:exist}
    There exists a representation of $\cS(A,\Phi)$ on a Hilbert space
    $H$ that satisfies $\langle a\rangle=tr\ a$ and $a^*$ is the
    adjoint of $a$ for $a\in\A$.
\end{prop}
\begin{proof}
    In \cite{MNoHyp} it is proven that for each charge $k$ monopole
    $(A,\Phi)$ there exists a holomorphic map
    $q:S^2_{\infty}\rightarrow\bbC\bbP^k$ with two key properties.  It
    determines and is determined by the spectral curve of $(A,\Phi)$
    and satisfies the statement of Proposition~\ref{th:orth}, and it
    determines and is determined by the boundary value $A_{\infty}$ of
    $(A,\Phi)$.  The curvature of $A_{\infty}$ is obtained as the
    pull-back of the Kahler form on $\bbC\bbP^k$ by $q$.

    As in the proof of Proposition~\ref{th:rephol}, use $|q(z)\rangle$
    to label a unit vector in the line $q(z)$, and $\langle q(z)|$ its
    conjugate transpose, so $|q(z)\rangle\langle q(z)|=R_z$ is
    well-defined.  We will prove that $R_z=R_z^*$ is the image of
    $P_z$ in a representation of $\A$ acting on $\bbC^{k+1}$
    satisfying $\langle P_{z_1}..P_{z_n}\rangle=tr\ 
    R_{z_1}..R_{z_n}=\langle q(z_1)|q(z_2)\rangle\langle 
    q(z_2)|q(z_3)\rangle..\langle q(z_n)|q(z_1)\rangle$. 
    Since $\langle a\rangle$ for any $a\in\A$ is obtained from
    derivatives and limits of such quantities, this is enough to show
    the representation satisfies (\ref{eq:rep}).
    
    The functions $\langle P_wP_z\rangle$ and $|\langle
    q(w)|q(z)\rangle|^2$ vanish to the same order on (an image under
    $w\mapsto\hat{w}$ of) the spectral curve of $(A,\Phi)$ and vanish
    nowhere else.  Thus, \[ \langle P_wP_z\rangle=\xi(w,z)|\langle
    q(w)|q(z)\rangle|^2\] for a real valued nowhere vanishing function
    $\xi(w,z)$.  Fix $q(w)$ and choose $q(z)$ so that $\langle
    q(w)|q(z)\rangle\in\bbR$ for each $z$.  Take the derivative of
    each side with respect to $\partial_{\bar{z}}$ so 
    \[ 2\lambda(w,z)\langle P_wP_z\rangle=(2\lambda(w,z)+
    \partial_{\bar{z}}\ln\xi(w,z))\xi(w,z)|\langle q(w)|q(z)\rangle|^2\] 
    since both $\langle P_wP_z\rangle$ and $|\langle q(w)|q(z)\rangle|^2$
    define $A_{\infty}=\lambda(z)d\bar{z}-\bar{\lambda}(z)dz$.  Hence
    \[\partial_{\bar{z}}\ln\xi(w,z)=0\] so $\xi(w,z)$ is constant.  It
    is identically $1$ since $\langle P_z^2\rangle=1=|\langle
    q(z)|q(z)\rangle|^2$.  

    Note that our assumption that $\langle P_wP_z\rangle$ and
    $|\langle q(w)|q(z)\rangle|^2$ define the same gauge for
    $A_{\infty}$ is unnecessary since if they differ by the gauge
    transformation
    \[\lambda(w,z)\mapsto\lambda(w,z)+i\partial_{\bar{z}}\theta(w,z)\]
    for a real-valued $\theta(w,z)$, then we are left with
    $\partial_{\bar{z}}\ln\xi(w,z)=-2i\partial_{\bar{z}}\theta(w,z)$
    in which case $\xi$ is harmonic and hence constant, thus
    $\theta\equiv 0$.
    
    The general case is proved analogously.  Again since we know the
    vanishing behaviour of the respective functions, we have \[\langle
    P_{z_1}..P_{z_n}\rangle=\xi(z_1,..z_n)\langle
    q(z_1)|q(z_2)\rangle\langle q(z_2)|q(z_3)\rangle..\langle
    q(z_n)|q(z_1)\rangle\] for a nowhere vanishing $\xi$.  Vary $z_1$
    and fix the other variables.  Choose $q(z_1)$ so that $\langle
    q(z_1)|q(z_2)\rangle\in\bbR$ for each $z_1$.  Then again 
    \[ 2\lambda(z_2,z_1)\langle P_{z_1}..P_{z_n}\rangle=(2\lambda(z_2,
    z_1)+(\partial_{\bar{z_1}}\ln\xi))\langle P_{z_1}..P_{z_n}\rangle\]
    and $\partial_{\bar{z_1}}\ln\xi(z_1,..,z_n)=0$.  Thus $\xi$ is 
    constant and it is 1 on the diagonal $z_i=z_1$, so it is 
    identically $1$.
\end{proof}

\begin{cor}  \label{th:pos}
$\langle a^*a\rangle\geq 0$ for any $a\in\A$, with equality precisely
when $a=0$.
\end{cor}
We have been unable to prove this property directly, requiring instead
Proposition~\ref{th:exist} and the positivity of the trace on the
product of a matrix with its adjoint.

Proposition~\ref{th:conenc} shows that
$F_{A_{\infty}}=-\langle[\partial_z,P_z][\partial_{\bar{z}},P_z]\rangle
dzd\bar{z}$ so a consequence of Corollary~\ref{th:pos} is the fact
that $F_{A_{\infty}}/2\pi i$ is non-negative with respect to the
orientation $idzd\bar{z}$.  Furthermore, we can also understand the
singularities of $q$ in terms of this curvature.  Since
$\partial_{\bar{z}}|q(z)\rangle=\lambda(z)|q(z)\rangle$, then $q$ is
singular at $z_0$ if and only if
$\partial_z|q(z)\rangle_{|z_0}=\mu|q(z_0)\rangle$ for some
$\mu\in\bbC$.  Now \[ 0=\partial_z\langle
q(z)|q(z)\rangle_{|z_0}=\langle
\partial_{\bar{z}}q(z)|q(z)\rangle_{|z_0}+\langle
q(z)|\partial_zq(z)\rangle_{|z_0}=\overline{\lambda}(z_0)+\mu\] thus
$[\partial_z,P_z]_{|z_0}=\partial_z|q(z)\rangle\langle
q(z)|_{|z_0}=(\overline{\lambda}(z_0)+\mu)|q(z_0)\rangle\langle
q(z_0)|=0$.  So by Corollary~\ref{th:pos}, $q$ has a singularity at
$z_0$ if and only if $F_{A_{\infty}}(z_0)=0$.

\section{Conclusion}
The important features of $\cS(A,\Phi)$ have thus far used the
bounded, real-valued 2-point function $\langle P_wP_z\rangle$.  The
3-point function was needed to prove some of the properties of
$\langle P_wP_z\rangle$.  Since the 2-point function determines the
algebra it might be that one need look not much further to the
$n$-point functions.  On the other hand, there are features of
$\cS(A,\Phi)$ that have yet to be understood and may require the
higher order functions.

(i) The existence of a finite-dimensional representation of
$\cS(A,\Phi)$ with expectation values of observables given by the
trace implies relations amongst the 4-point functions.  More
precisely, for a charge $k$ monopole, choose a generic set of points
$\{ z_i|i=0,..,N\}$ (where $N$ is the dimension of the span of the
image of $q(z)$, so $N=k$ if $q$ is ``full'') and set $P_i=P_{z_i}$. 
Then the finite dimensional representation allows any $P_w$ to be
expressed as $\alpha^{ij}(w)P_iP_j$ (sum repeated indices) where the
$\alpha^{ij}(w)$ are determined via $\langle
P_wP_kP_l\rangle=\alpha^{ij}(w)\langle P_iP_jP_kP_l\rangle$.  Set
$g_{ijkl}=\langle P_iP_jP_kP_l\rangle$.  Then (for generic choice $\{
z_i|i=0,..,N\}$) there exists an ``inverse'' $g^{ijkl}$ satisfying
$g^{ijkl}g_{klmn}=\delta_{im}\delta_{jn}$, so
$\alpha^{ij}(w)=g^{ijkl}\langle P_wP_kP_l\rangle$.  Then, \[\langle
P_wP_z\rangle=g^{ijkl}\langle P_wP_kP_l\rangle\langle
P_zP_iP_j\rangle.\] If we multiply both sides by the ``determinant''
of $g_{ijkl}$ then the relation holds for all sets $\{
z_i|i=0,..,N\}$, and not just generic sets.  It would be more
satisfying to be able to prove the relations directly and use this to
get the representation.

(ii) It would be interesting to recognise the mass of the monopole in
$\cS(A,\Phi)$.  The mass is encoded in the spectral curve but it is
difficult to extract.

(iii) Since $\cS(A,\Phi)$ brings the spectral curve of $(A,\Phi)$ and
the connection on the conformal boundary two-sphere closer together,
one might hope to understand both the metrics of Austin and Braam
\cite{ABrBou} and Hitchin \cite{HitNew} from a similar perspective.

(iv) One can take finite-dimensional subalgebras of $\cS(A,\Phi)$ to
possibly uncover further structure.  In the case $k=2$, define
$\cS_w(A,\Phi)\subset\cS(A,\Phi)$ to be the sub-algebra generated by
$P_1(w)=P_{z_1}$ and $P_2(w)=P_{z_2}$ where $P_wP_{z_i}=0$.  This is a
finite-dimensional algebra, generated as a vector space by $P_1(w)$,
$P_2(w)$, $P_1(w)P_2(w)$ and $P_2(w)P_1(w)$.  The algebra
$\cS_w(A,\Phi)$ actually depends on a point in the spectral curve of
the monopole, since the elements $P_1(w)$ and $P_2(w)$ are ordered.
Each point of the spectral curve stores information such as the
structure coefficients of the finitely generated sub-algebras.  The
interaction of the sub-algebras at different points also encodes
information.  Another interesting class of sub-algebras parametrised
by the spectral curve arises from monopoles invariant under a
$\bbZ_k$-action.  For any point of the spectral curve, take its orbit
of $k$ (ordered) points and take the forward endpoints
$w_1,w_2,..,w_k$ say, defining the subalgebra to be generated by
$P_i=P_{w_i}$.  Algebras depending on a spectral parameter arise in
many parts of mathematics.  It would be interesting to understand how
these families of algebras depending on points of the spectral curve fit
into other constructions.

(v) The algebra $\cS(A,\Phi)$ of an $SU(2)$ hyperbolic monopole
generalises to any gauge group.  In such a case, the scattering
equations (\ref{eq:scat}) admit solutions with various rates of decay.
To each point $z\in S^2_{\infty}$ we associate finitely many operators, one
for each level of decay of solutions of the scattering equation, with
given relations.  The $n$-point functions are obtained from pairing
solutions of the scattering equations with specified decay in each
direction.  For higher rank Lie groups, just as the operators $P_z$
define one-dimensional subspaces of a very large vector space to give
a holomorphic map $q:S^2_{\infty}\rightarrow\bbC\bbP^k$, the finitely many
operators associated to $z\in S^2_{\infty}$ will define a flag inside a very
large vector space with a corresponding holomorphic map.  The
dimension of the vector space will be determined by the charge of the
monopole, as in Proposition~\ref{th:rephol}.

\end{document}